\documentclass[twocolumn]{aastex631}

\usepackage{amsmath}
\usepackage{rotating}
\usepackage{tabularx}
\usepackage{float}

\shorttitle{Solar Neighborhood Cloud Catalog}
\shortauthors{Cahlon et al.}
\graphicspath{{./}{figures/}}
\DeclareUnicodeCharacter{2212}{-}
\begin{document}
\title{A Parsec-Scale Catalog of Molecular Clouds in the Solar Neighborhood Based on 3D Dust Mapping: Implications for the Mass-Size Relation}

\author[0000-0002-7072-5627]{Shlomo Cahlon}
\affiliation{Center for Astrophysics $\mid$ Harvard \& Smithsonian, 
60 Garden St., Cambridge, MA, USA 02138}

\author[0000-0002-2250-730X]{Catherine Zucker}
\affiliation{Center for Astrophysics $\mid$ Harvard \& Smithsonian, 
60 Garden St., Cambridge, MA, USA 02138}
\affiliation{Space Telescope Science Institute, 3700 San Martin Drive, Baltimore, MD 21218, USA}

\author[0000-0003-1312-0477]{Alyssa Goodman}
\affiliation{Center for Astrophysics $\mid$ Harvard \& Smithsonian, 
60 Garden St., Cambridge, MA, USA 02138}

\author[0000-0002-4658-7017]{Charles Lada}
\affiliation{Center for Astrophysics $\mid$ Harvard \& Smithsonian, 
60 Garden St., Cambridge, MA, USA 02138}

\author[0000-0002-4355-0921]{Jo\~{a}o Alves}
\affiliation{University of Vienna, Department of Astrophysics, T{\"u}rkenschanzstra{\ss}e 17, 1180 Vienna, Austria}

\correspondingauthor{Shlomo T. Z. Cahlon}
\email{stz.cahlon@gmail.com}



\begin{abstract}

We dendrogram the \cite{Leike} 3D dust map, leveraging its  $\sim 1$ pc spatial resolution to produce a uniform catalog of molecular clouds in the solar neighborhood. Using accurate distances, we measure the properties of 65 clouds in true 3D space, eliminating much of the  uncertainty in mass, size, and density. Clouds in the catalog contain a total of $1.1 \times 10^5 \; \rm M_{\sun}$, span distances of $116-440$ pc, and include a dozen well-studied clouds in the literature. In addition to deriving cloud properties in 3D volume density space, we create 2D dust extinction maps from the 3D data by projecting the 3D clouds onto a 2D ``Sky" view. We measure the properties of the 2D clouds separately from the 3D clouds. We compare the scaling relation between the masses and sizes of clouds following \cite{1981Larson}. We find that our 2D projected mass-size relation, $M \propto r^{2.1}$, agrees with Larson’s Third Relation, but our 3D derived properties lead to a scaling relation of about one order larger: $M \propto r^{2.9}$. Validating predictions from theory and numerical simulations, our results indicate that the mass-size relation is sensitive to whether column or volume density is used to define clouds, since mass scales with area in 2D ($M \propto r^{2}$) and with volume in 3D ($M \propto r^{3}$). Our results imply a roughly constant column and volume density in 2D and 3D, respectively, for molecular clouds, as would be expected for clouds where the lower density, larger volume-filling gas dominates the cloud mass budget. 

\end{abstract}



\section{Introduction} \label{sec:intro}
Star formation takes place in molecular clouds, which are associated with the densest and coldest phase of the interstellar medium (ISM). Studying the properties of molecular clouds has thus long been the focus of star formation research, as the structure of these clouds has a direct impact on the location, number, size, and mass of newly formed stars \citep{2008Rosolowsky}.  

Maps of the extinction or emission from dust trace out the interstellar medium (ISM) in ``position-position" or ``{\it p-p}" space, on the 2D plane of the sky \citep{Lombardi_2009, Lada_2009}. Spectral-line observations of the ISM can add a third dimension, owing to the Doppler effect, which allows for conversion of wavelength or frequency to velocity. The resulting so-called ``postion-position-velocity" or  ``\textit{p-p-v}" cubes, can be analyzed as 2D maps (integrating over velocity) or as pseudo-3D maps, where velocity is treated as a non-spatial third dimension. 

Catalogs of molecular clouds have previously been derived using both \textit{p-p} and \textit{p-p-v} data. \cite{2016_Rice} use the dendrogram technique \citep{2008Rosolowsky} to extract and analyze molecular clouds from the CO \textit{p-p-v} survey of \citet{Dame_2001}, identifying over a thousand clouds across the full Galactic plane. \cite{Miville_Desch_nes_2016} apply a hierarchical cluster identification method to a Gaussian decomposition of \citet{Dame_2001} and produce a catalog of 8,107 clouds covering the entire Galactic plane. Using 2D extinction maps derived from the NICEST color excess method \citep{Lombardi_2009}, \cite{Dobashi_2011} identify over 7,000 dark clouds in the Galactic plane using a fixed extinction threshold. 

Numerical simulations show that projection effects intrinsic to \textit{p-p} and \textit{p-p-v} space impact the study of cloud structures \citep[e.g.][]{2010ShettyRosoGoodman, 2002parades_mac}. Comparing simulated {\it p-p-p-} and {\it p-p-v-}derived clouds' overlap,  \citet{2013Beaumont} find that studying clouds in \textit{p-p-v} space (rather than in true physical 3D space) can induce approximately 40\% scatter in their masses, sizes, and velocity dispersions. Moreover, \cite{2013Beaumont} demonstrate that many \textit{p-p-v} structures can be fictitious, especially in ``crowded" regions. Thus, accurate estimates of cloud properties depend critically on studying clouds in \textit{position-position-position} space, which requires knowledge of clouds' distances. 

In the past few years, distance estimates to molecular clouds have improved dramatically. Using so-called 3D dust mapping, \cite{2014_schalafly(finkbeinerGroup)} produce one of the first uniform catalogs of accurate distances to nearby molecular clouds, with typical distance uncertainties of $\approx 10\%$. Specifically,  \cite{2014_schalafly(finkbeinerGroup)} use multi-band photometry from Pan-STARRS1 \citep{PS1} to infer self-consistent distances and extinctions for a large number of stars across the solar neighborhood, the key ingredients necessary for constructing a 3D dust map \citep[see also][]{Green_2015}.  The advent of the \textit{Gaia} mission \citep{Gaia_2016}, and especially the results from its second and third data releases, \textit{Gaia} DR2 and DR3 \citep{2018_gaia, 2021_gaia}, has made it possible to construct ever-more-accurate 3D-dust-based distances to clouds, owing to stellar parallax measurements for millions of stars in the solar vicinity.  \cite{2019zucker} utilize the \textit{Gaia} DR2 data release to produce an accurate catalog of distance estimates to molecular clouds, with uncertainties on the order of $5\%-6\%$ \citep[see also][]{2020zucker, yan2019}

Building on the accurate distances enabled in the \textit{Gaia} era, there have only been two molecular cloud catalogs based on true three-dimensional ``{\it p-p-p}" data obtained from 3D dust mapping, as presented in \citet{Chen_2020} and \citet{Dharmawardena_2022}.  \citet{Chen_2020} obtain a catalog of 567 molecular clouds using the 3D dust map of \citet{Chen_2019}. However, the molecular clouds are typically resolved in distance on $\approx$ hundreds of parsec scales, so key cloud properties, including the sizes of clouds, are still derived using 2D projections. \citet{Dharmawardena_2022} also derive a catalog of molecular cloud properties towards sixteen complexes within $\rm 1-2 \; kpc$ from the Sun using their 3D dust mapping algorithm \texttt{DUSTRIBUTION}, which leverages stellar distance and extinction estimates from \citet{Fouesneau_2022}. Applying the \texttt{astrodendro} package \citep{astrodendro} to 3D dust cutouts around each complex, \citet{Dharmawardena_2022} obtain estimates of e.g. the volume, mass, and density for each cloud and its myriad of substructure in ``{\it p-p-p}" space \citep[see also][]{Dharmawardena_2021}.

However, one of the highest resolution 3D dust maps over appreciable volumes of the solar neighborhood is the \citet{Leike} map, which traces the structure of the local interstellar medium at $\sim 1$ pc distance resolution. Leveraging distance and extinction estimates from the StarHorse catalog \citep{Anders_2019},  \cite{Leike} utilize a combination of Gaussian Processes and Information Field Theory to produce a highly resolved 3D dust map that charts molecular clouds out to a distance of $d \approx 400$ pc with distance uncertainties lower than 1\%. Such accurate distance uncertainties enable the extraction and characterization of molecular clouds in true 3D {\it p-p-p} space.

In this work, we dendrogram the \cite{Leike} 3D dust map and uniformly analyze the properties of resolved molecular clouds derived in {\it p-p-p} space. We produce a catalog of 65 distinct local molecular clouds, including a dozen well-studied clouds in the literature, and compare our results to extant literature derived primarily from \textit{p-p-v} and \textit{p-p} space. In \S \ref{sec:data} we present the \cite{Leike} data used to create the catalog. In \S \ref{sec:method} we present the dendrogram technique applied to the data to derive the properties of our molecular clouds in real 3D space. We then describe how we project our data into 2D space following \citet{Zucker_2021}, in order to measure the 2D properties of clouds. In \S \ref{sec:results} we summarize our cloud property results and characterize Larson's mass-size relation in both 2D and 3D space. In \S \ref{sec:discussion} we hypothesize what could be driving differences in the 3D- and 2D- derived mass-size relations, and discuss our mass-size results in the context of existing literature, including previous exploration of the mass-size relation using analytic theory and mathematical modeling \citep[c.f.][]{Ballesteros_Paredes_2012}. Finally, we conclude in \S \ref{sec:conclusions}

\section{Data} \label{sec:data}

 \cite{Leike} reconstruct the 3D dust distribution in a Heliocentric Galactic Cartesian reference frame out to a distance of $\approx 400$ pc ($-370\; {\rm pc} < xy < 370 \; {\rm pc}, -270 \; {\rm pc} < z < 270 \; {\rm pc}$). This distance range includes about a dozen well-studied star-forming regions, including Taurus, Perseus, and Orion. We convert from the native units of the \citet{Leike} map (optical depth in the $Gaia$ $G$ band per 1 pc) to volume density of hydrogen nuclei ($n_{\rm H}$) following \citet{Zucker_2021} and \citet{Bialy_2021}. We derive all results in this work using the total volume density of hydrogen nuclei, including contributions from both atomic and molecular hydrogen gas.

\begin{figure*} \label{3-panal-size-mass}
    \centering
    \includegraphics[width=150mm]{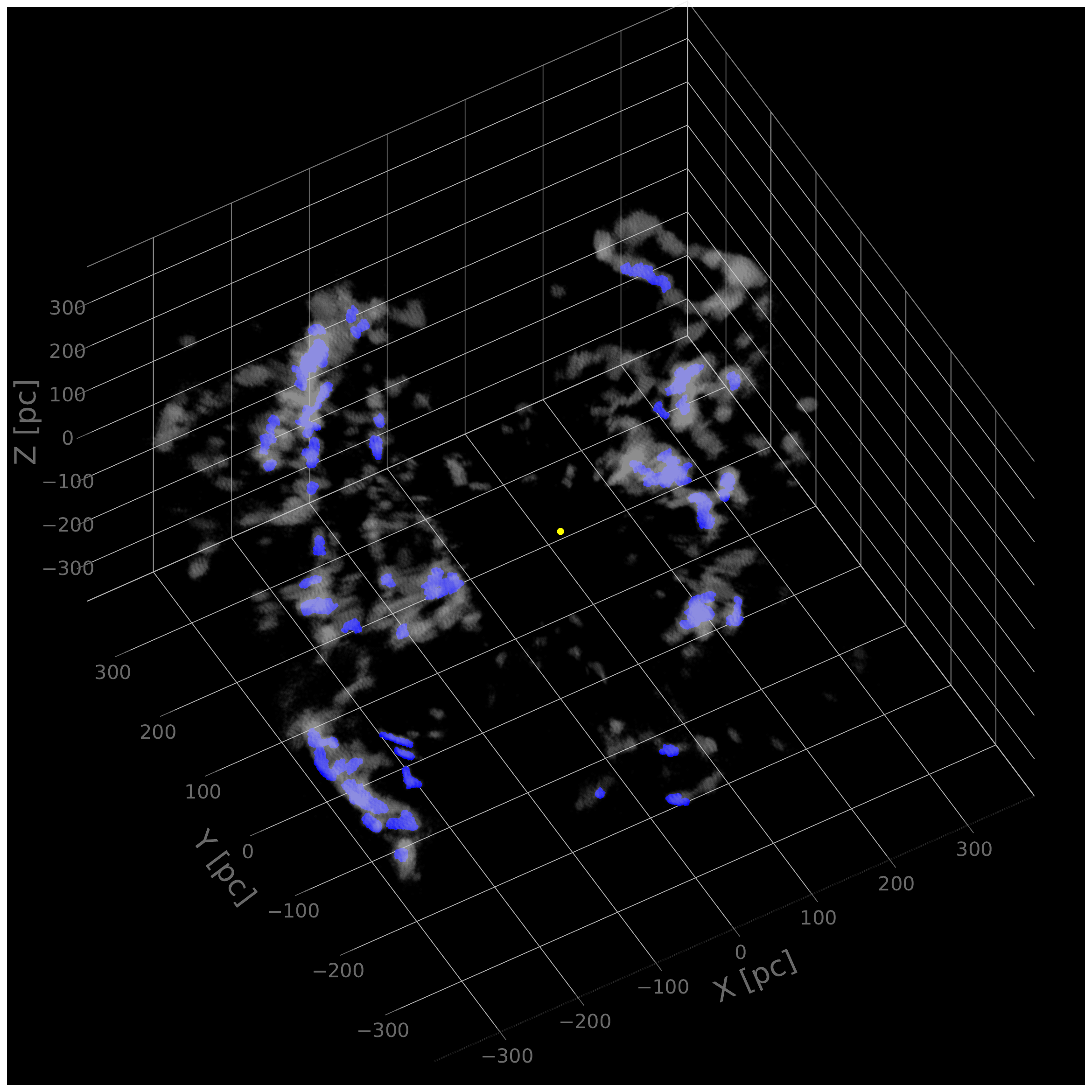}
    \caption{ 3D spatial map of the local interstellar medium, showing all gas with density $n > 5\; \rm cm^{-3}$ (gray) and clouds identified in this work (blue). The Sun is shown at center in yellow. An interactive version of this figure is available \href{https://faun.rc.fas.harvard.edu/czucker/Paper_Figures/3D\_Cloud\_Catalog\_Cahlon\_2023.html}{here}.}
    \label{fig:coloredClouds}
\end{figure*}

\section{Methods} \label{sec:method}
\subsection{Generating 3D Dendrogram} \label{subsec:dendro_params}

After converting the 3D dust map of \cite{Leike} to total volume density of hydrogen nuclei, we segment the \cite{Leike} 3D dust map into a set of molecular cloud features and measure their properties using the dendrogram algorithm. To do so, we build upon the existing functionality for dendrogramming 3D \textit{p-p-v} data in the \texttt{astrodendro} package. Abstractly, the dendrogram  algorithm constructs a tree starting from the highest density point in N-dimensional density (volume density in this work) data, moving to the next largest value and connecting along isosurfaces of constant density. A leaf is defined to be a feature without any descendants. Each time a local maximum point is found (i.e. a leaf), the algorithm determines, based on neighboring maxima and the behavior of the contour levels between maxima, whether to join the pixel to an existing structure, or to create a new structure. Once a local minimum point between the two structures is found, it is classified as a branch that connects the two structures. Iterations of this procedure will eventually either merge all values into a single tree or create multiple trees. Moreover, once the data are contoured with levels, the algorithm searches through every contour level, starting from the top, and records how many local maxima are above each contour level. When the surface around two local maxima merge together, that density level is recorded as a branch. If more than two local maxima merge together between two successive contour levels, the algorithm will continue to search with better tuned contour levels such that every merger includes up to two leaves. The dendrogram algorithm depends on three user-defined parameters set in the \texttt{astrodendro} package, $n_{min}$, $\Delta_{n}$, and $\#_{voxels}$:

\begin{enumerate}
  \item $n_{min}$: the minimum absolute volume density threshold for a structure to be included as part of the dendrogram.
  \item $\Delta_n$: how significant a leaf must be in order to be considered an independent entity. The significance is measured from the difference between its peak density and the density value at which it is being merged into the tree.
  \item $\#_{voxels}$: the minimum number of voxels needed for a leaf to be considered an independent entity. If a leaf is about to be joined onto a branch or another leaf, the algorithm checks the leaf's number of voxels. If the leaf's number of voxels is lower than $\#_{voxels}$, the algorithm combines it with the branch or leaf it is being merged with, so that it is no longer considered a separate entity.
\end{enumerate}

The dendrogram is most sensitive to changes in $n_{min}$, as this value determines the minimum volume density value required to classify local maxima as meaningful. Setting a high $n_{min}$ can lead to multiple isolated single-leaf trees, especially when dealing with 3D dust maps. To determine what values to adopt for these parameters, we created a set of different combinations of values and compared the subsequent masses computed using these values to a benchmark sample of about a dozen well-studied clouds in \textit{p-p} space from \cite{Lada_2010} and \textit{p-p-p} space from \citet{Zucker_2021}. We tailored the parameters to reproduce similar results to the bench-marked cloud samples. After testing multiple values for the parameters, we ultimately settled on a fixed threshold of: $n_{min}=25 \; \rm cm^{−3}$, $\Delta_{n} = 25 \; \rm cm^{-3}$, $\#_{voxels} = 150$ voxels. The result of this procedure is a hierarchy of cloud emission, where each structure in the dendrogram corresponds to a contiguous, resolved feature in 3D Heliocentric Galactic Cartesian $xyz$ coordinates, bounded by a surface of constant volume density. In the next section, we filter the dendrogram to extract clouds and their properties. 

\begin{figure*} 
    \centering
    \includegraphics[width=1.0\textwidth]{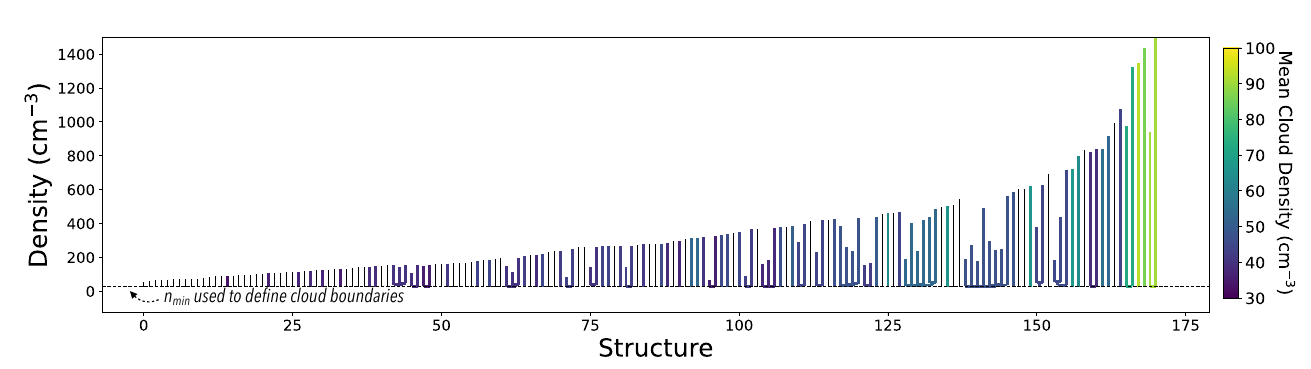}
    \caption{Dendrogram of the \citet{Leike} 3D dust map. Structures that survive the cloud filtering scheme summarized in \S \ref{sec: filtering} are color-coded by their mean density ($n_{average}$ in the context of Table \ref{tab:table_1}), with filtered out structures shown in black (totaling $3.7 \times 10^4 \; \rm M_{\sun}$). A constant isosurface density of $n_{min} = 25 \; \rm cm^{-3}$ (dashed horizontal line) is used to define cloud boundaries and will pre-ordain the narrow range of mean cloud volume densities ($33 \; {\rm cm^{-3}} \leq n_{average} \leq 92 \; {\rm cm^{-3}}$) seen in our sample. \label{fig:full_dendrogram}} 
\end{figure*}

\subsubsection{Filtering 3D Features} \label{sec: filtering}
After generating a hierarchy of cloud structure using the dendrogram approach, we need to convert the dendrogram tree into a meaningful set of molecular cloud features for analysis. In order to avoid both spurious features (with relatively low mass) and double counting of nested clouds (i.e. counting a branch and its leaves as separate structures), we introduce a filtering scheme.

Our filtering scheme is based on the adoption of a minimum cloud mass, $M_{min}$, and cloud radius, $r_{min}$, required for a feature to be included in the catalog. The mass is computed for all features as described in \S \ref{measuring3D} below. We define $M_{min}=500 \; \rm M_\odot$ and $r_{min}=2 \; \rm pc$. By using this definition, the algorithm only includes structures that correspond to trunks in the dendrogram that are above $M_{min}$ and $r_{min}$. All other structures are removed to avoid counting nested clouds, which by definition are not trunks. 

Given that most dendrograms computed from 3D dust maps are composed of an ensemble of single structure trees, retaining only the trunk feature means retaining the structures defined by isosurfaces near $n_{min} = 25$ cm$^{-3}$. The filtered molecular cloud features extracted using this approach are highlighted and overlaid on the underlying \citet{Leike} 3D dust map in Figure \ref{fig:coloredClouds}. The implication of this filtering method is that we are removing most of the cloud hierarchy and limiting the extracted clouds to a narrow range in mean cloud density. In theory, we could achieve a similar catalog by thresholding the \citet{Leike} volume above a density of $n_{min} = 25$ cm$^{-3}$. However, we choose to dendrogram both to utilize the existing functionality of the \texttt{astrodendro} package \citep{astrodendro} and to enable further follow-up studies of the full hierarchy. 

In Figure \ref{fig:full_dendrogram}, we show the full dendrogram, and highlight the cloud features that survive filtering. The combined mass of the filtered out structures is $3.7 \times 10^4 \; \rm M_{\sun}$ and are shown in black in Figure \ref{fig:full_dendrogram}. The remaining, surviving clouds are color-coded by the mean density, with the clouds possessing mean densities between $n_{average}=33-92 \; \rm cm^{-3}$. As we will discuss further in \S \ref{sec:discussion}, the narrow range in density of extracted clouds will pre-ordain the mass-size results we obtain in \S \ref{sec:results} due to the large filling factor of low density gas near the chosen $n_{min}$ \citep[see e.g. further discussion in][]{Beaumont_2012}. 

 Finally, we also recognize that a relatively low volume density isosurface is defining the 3D cloud boundaries. However, it is not possible to define clouds using significantly higher volume density thresholds because the \citet{Leike} 3D dust map is not sensitive to the highest volume density regions within molecular clouds, due to the map's reliance on optical stellar photometry and astrometry from \textit{Gaia} \citep{2018_gaia}. Despite not being sensitive to very high volume densities, \citet{Zucker_2021} show that the \citet{Leike} is still recovering cloud properties based on 2D integrated approaches, as determined for a benchmark sample of well-studied clouds. Thus, the low isosurface levels defining cloud boundaries should not have a significant effect on our mass results.

\subsubsection{Measuring Cloud Properties in 3D} \label{measuring3D}

To calculate the properties of clouds in \textit{p-p-p} space, we again build on existing infrastructure for computing cloud properties in \textit{p-p} and \textit{p-p-v} space using the \texttt{astrodendro} package \citep{astrodendro}. In order to calculate cloud properties, we take as input the dendrogrammed cloud structure, where each cloud structure consists of a set of contiguous volume density voxels bounded by a surface of constant volume density. We first calculate physical properties for all structures, and subsequently define the final cloud catalog through filtering as explained in \S \ref{sec: filtering}. Our catalog includes the following properties for every cloud (see Table \ref{tab:table_1}). \footnote{We adapted and extended a version of the \texttt{astrodendro} package written by Dario Colombo and Ana-Duarte Cabral. The software is available here: \url{https://github.com/dendrograms/astrodendro/pull/147/files}}

\begin{enumerate} \label{definingParameters}

  \item $V$ (pc$^3$): Exact volume of the structure in \textit{p-p-p} space  
  \item $r$ (pc): Equivalent radius of the sphere occupying the same volume as the volume exact $V$
  \item $x, y, z$ (pc): Central $x$, $y$, and $z$  position of the cloud in Heliocentric Galactic Cartesian coordinates
   \item $l$ ({\scriptsize $^\circ$}), $b$ ({\scriptsize $^\circ$}), and $d$ (pc). The cloud's Galactic coordinates (longitude, latitude, and distance), computed from the mean $x$, $y$, and $z$  position of the cloud in Heliocentric Galactic Cartesian coordinates
  \item $M$ (M$_{\odot}$): the mass of the cloud, calculated as follows: \begin{equation} \label{equation_massToDensity}
        M = \Sigma \; dM_i = \; \Sigma \; 1.37\;m_p \times n_i \times dV_i
  \end{equation}
 The total mass is the sum of the mass in each individual voxel $dM_i$, computed by multiplying the volume density in the $i$th voxel ($n_i$) by the mean molecular weight of hydrogen ($1.37 \times m_p$, correcting for the helium abundance) and  its volume (1 pc$^3$)
  \item $n_{peak}$ (cm$^{-3}$): maximum volume density within the cloud
  \item $A$ (pc$^2$): surface area of the cloud, calculated by assuming a spherical geometry with a radius of $r$ and determining the cross-sectional area ($\pi r^2$)
  \item $\Sigma$ ($\rm \frac{M_\odot}{pc^2}$): the mass surface density of the cloud, given as the cloud's mass divided by its surface area
  \item $n_{average}$ (cm$^{-3}$): average volume density of the cloud
\end{enumerate}

\subsection{2D Methods}

\subsubsection{Converting 3D Dust Data into 2D Extinction Maps} \label{2d_converting_section}

Once a catalog of cloud features is defined and characterized in 3D, we create a 2D projected map of each 3D cloud following the methodology of \citet{Zucker_2021} (see their \S 3.3 for full details).  Briefly, that work uses the \texttt{yt} package \citep{Turk_2011} to integrate 3D volume density cubes (containing the 3D cloud of interest) along the line of sight and produce 2D maps of the total hydrogen column density. For each cloud, we obtain a 3D volume density sub-region suitable for projection by extracting a cutout of the \citet{Leike} 3D dust map using the minimum and maximum extent of the cloud boundaries along $x$, $y$, and $z$. Once we obtain the projected 2D column density maps, we convert from total hydrogen column density to visual extinction in the $K$ band using a relation from \citet{Lada_2009} of $\frac{N({\rm HI}) + 2N({\rm H_{\rm 2}})}{A_K} = 1.67 \times 10^{22}  \rm  \; cm^{-2} \; mag^{-1}$ to produce a map of $K$-band extinction, $A_K$. Converting to $A_K$ allows us to compare to previous 2D cloud catalogs built on similar maps of integrated dust extinction from \citet{Lada_2009}.

To analyze the 2D maps, we extract clouds on the plane of the sky using the existing \textit{p-p} dendramming functionality of the \texttt{astrodendro} package. In order to understand how 3D clouds map to 2D projected space we use \citet{Zucker_2021} as a guide, who analyze the 3D cartesian space ($x,y,z$) and Galactic ($l,b,d$) coordinates of a sample of famous nearby clouds. Identifying the relevant plane of the sky features from the 3D projected famous cloud data, we then use \cite{Lada_2009} as a benchmark to determine the optimal $A_{K_{{min}}}$, $\Delta_{A_K}$, and $\#_{pixels}$ parameters (where $A_{K_{{min}}}$, $\Delta_{A_{K}}$, and $\#_{pixels}$ are the 2D analogs of the $n_{min}$, $\Delta_{n}$, and $\#_{voxels}$ 3D input parameters described in \S \ref{subsec:dendro_params}) for computing the dendrogram, with the goal of obtaining similar cloud sizes as derived in \citet{Lada_2009}. With the intention of also having a single 2D structure representing each 3D cloud feature, we settle on $A_{K_{{min}}} = 0.05$ mag, $\Delta_{A_{K}} = 0.05$ mag, and $\#_{pixels} = 300$ pixels as the dendrogram input parameters. 

\begin{figure}
    \centering
    \includegraphics[width=80mm]{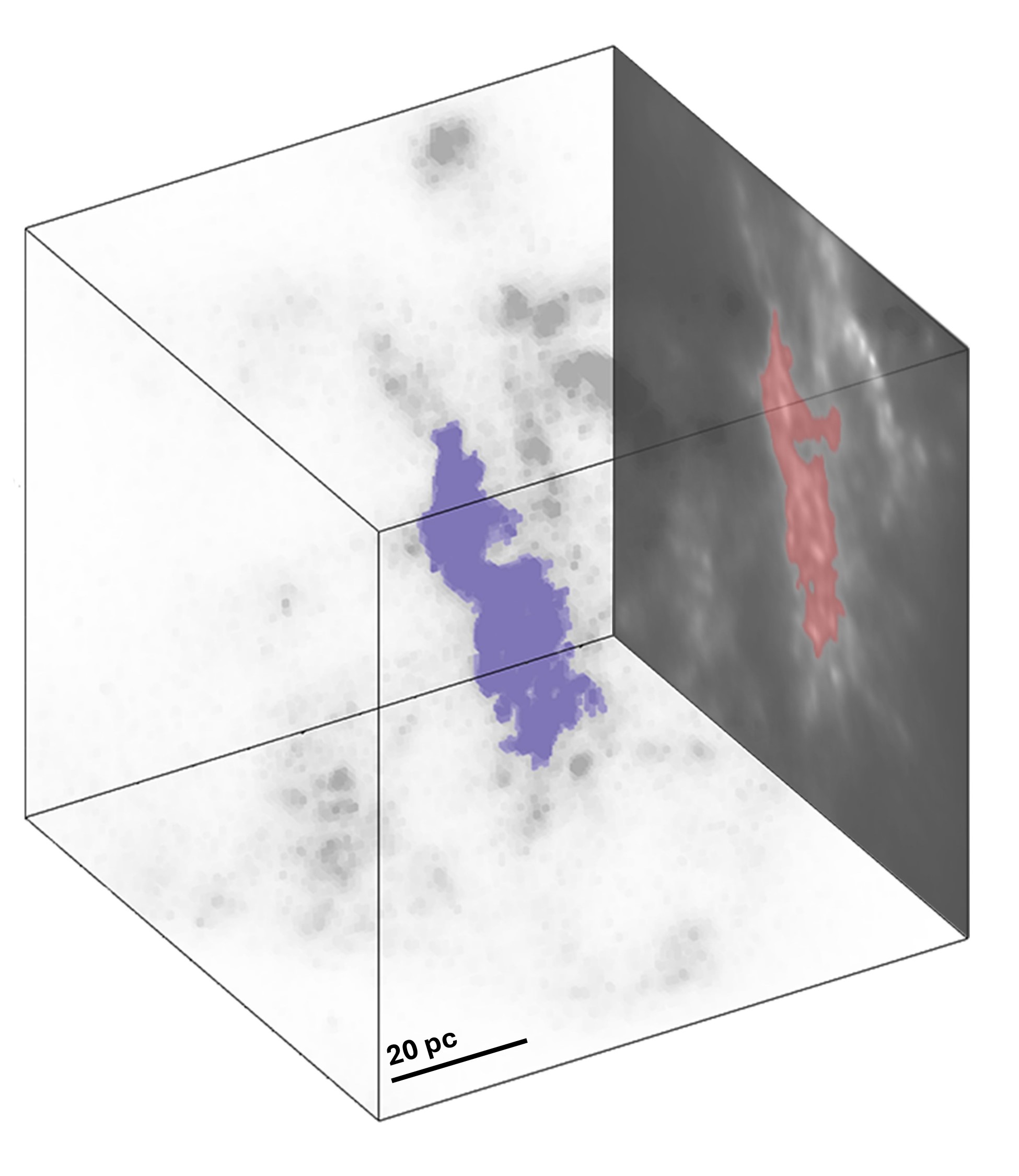}
    \caption{Comparison between the 3D and projected 2D dust data for the Perseus Molecular Cloud. The feature corresponding to the Perseus Molecular Cloud in 3D volume density space is shown in blue. The 3D dust has been projected onto the plane of the sky and shown via an $A_K$ extinction map in the background, where the 2D feature corresponding to the Perseus Molecular Cloud is shown in red.}
    \label{fig:3d_2d}
\end{figure}

In Figure \ref{fig:3d_2d}, we show an example of the 3D to 2D projection for a single feature in the catalog (the Perseus Molecular Cloud). We emphasize that due to the imperfect mapping from 3D to 2D space --- stemming from the complex geometries of individual clouds \citep{Zucker_2021} --- a few 3D features do not have a 2D counterpart, largely because they were sub-divided into multiple components and failed to produce a cloud feature with the same mass and/or size minima adopted for the 3D catalog. Our goal in this work is not to measure the most accurate 2D-based properties of molecular clouds. Rather, we seek to understand how defining features in 3D versus 2D projected space (given the same underlying 3D data) affects cloud properties in aggregate. 

Nevertheless, for clouds that have a 2D counterpart meeting these criteria (61/65 clouds, or $\approx 94\%$ of the 3D sample), the morphological matching between 3D and 2D cloud shapes is clear. We show and discuss the correspondence between 3D and 2D further in Appendix \S \ref{sec:appendix_b}

\subsubsection{Calculating Properties for 2D Projected Structures} \label{measuring2D}

After creating the dendrogram based on the $A_K$ maps, we filter the features in our 2D catalog by implementing the same minimum mass threshold ($M_{min}=500 \; \rm M_\odot$) and radius threshold ($r_{min}=2 \; \rm pc$) as our 3D data. In 2D space we measure the following properties:

\begin{enumerate}

\item $l$ ({\scriptsize $^\circ$}): Central longitude of the cloud 
\item $b$ ({\scriptsize $^\circ$}): Central latitude of the cloud 
\item $d$ (pc): The distance of the cloud\footnote{Retrieved from the 3D cloud structure before projecting.}
\item $A$ (pc$^2$): Exact surface area of the structure in \textit{p-p} space
\item $M$ (M$_{\odot}$): Mass of the cloud, derived using the following mass surface density relation from \citet{Zari_2016} and the distance of the cloud originally detected in 3D: \begin{equation}
    \frac{\Sigma_{\rm gas}}{A_K} = \mu m_p \beta_K = 183 \; {\rm M_\odot} \; \rm pc^{-2} \; mag^{-1}    
\end{equation}
\item $r$ (pc): Radius calculated using the exact area of the structure in \textit{p-p} space $\bigl(\sqrt\frac{A}{\pi}\bigr)$ assuming a spherical geometry
\item $\Sigma$ ($\rm \frac{M_\odot}{pc^2}$): the mass surface density of the cloud, given as the cloud's mass divided by its surface area
\end{enumerate}

\section{Results} \label{sec:results}

\subsection{Summary of 3D and 2D Cloud Properties}

In Table \ref{tab:table_1}, we present a summary of the properties of molecular clouds derived in 3D space following \S \ref{measuring3D}. A machine readable version of Table \ref{tab:table_1} and its associated \texttt{astrodendro} dendrogram file is available online at the Harvard Dataverse (\href{https://doi.org/10.7910/DVN/BFYDG8}{DOI:10.7910/DVN/BFYDG8}). The 3D catalog contains a total mass of $1.1 \times 10^{5} \; \rm M_{\odot}$ across the 65 cloud features identified in 3D volume density space, with an average cloud mass of $M = 1.7 \times 10^{3} \; \rm M_\odot$. The distance range of the clouds spans $d = 116 - 440$ pc. The typical average density of clouds in the 3D catalog is $n_{\rm average} = 47 \; \rm cm^{-3}$, while the typical peak density is about an order of magnitude higher ($n_{\rm peak} = 414 \; \rm cm^{-3}$). We find an average cloud volume of $V = 1220 \; \rm pc^{3}$ and an average equivalent radius assuming a spherical geometry of $r = 6$ pc, though we emphasize that some of the clouds show more complex, extended geometries. While every 3D cloud feature is assigned a unique identifier, the catalog includes a number of well-studied clouds in the literature, including Perseus, Taurus, Lupus, Chamaeleon, Cepheus, and the Orion complex (Orion A, Orion B, and $\lambda$ Orion) which have been denoted as such in a separate column in Table \ref{tab:table_1} to aid comparison with existing studies. 

In Table \ref{tab:table_2} we present the corresponding catalog of 2D cloud properties derived from the projected 3D data following \S \ref{measuring2D}. A machine readable version of Table \ref{tab:table_2} is likewise available online at the Harvard Dataverse (\href{https://doi.org/10.7910/DVN/BFYDG8}{DOI:10.7910/DVN/BFYDG8}). The 2D catalog contains a total mass of $2 \times 10^{5} \; \rm M_{\odot}$ across the 61 cloud features derived from the projection of the 3D cloud data. One of these 61 clouds, feature 7, is broken into two components, leading to a total of 62 clouds in Table \ref{tab:table_2}. Moreover, in Table \ref{tab:table_2} we use the same cloud identifiers as Table \ref{tab:table_1}, which specifies how each 2D cloud feature maps to its 3D counterpart. In the case a cloud in Table \ref{tab:table_1} has no corresponding cloud identifier in Table \ref{tab:table_2}, the cloud was filtered for not meeting our minimum mass threshold of 500 $\rm M_\odot$. The typical radius of clouds in the 2D catalog is marginally larger than the 3D catalog, averaging $r = 7$ pc, and the average cloud mass is about a factor of two higher, at $M = 3.2 \times 10^{3} \; \rm M_\odot$. 

Considering the ensemble of clouds, the total mass of the entire 2D-derived catalog is approximately $\approx 1.9 \times$ higher than the total mass of the 3D-derived catalog. The discrepancy suggests that projecting 3D gas density into 2D can alter the perceived shape of molecular clouds enough to bear significantly on their derived mass. This effect likely stems from the complications of projecting a non-spherical 3D cloud geometry onto the plane of the sky, resulting in a different cloud boundary definition. However, diffuse emission in the vicinity of the cloud also likely plays a major role. 

As a testament to the impact of diffuse intervening gas, recall that we create 2D dust maps by projecting 3D dust cutouts which were extracted using a bounding box corresponding to the the minimum and maximum extents of the 3D cloud features in $xyz$ space. Over the sample of 3D cutouts, we compute the ratio of the mass inside the 3D features to the mass outside the feature but within the bounding box used for 2D projections. We find that, over the full sample, there is $2\times$ as much mass outside the 3D dendrogrammed features as within them. Thus, if even half of this excess mass in the vicinity of each cloud is incorporated into the 2D cloud definitions, this contamination would be enough to account for the discrepancy in total mass between the 3D- and 2D-derived catalogs that we observe. 

\subsection{Fitting the Mass-Size Relation}
We use the masses and sizes in Table \ref{tab:table_1} and \ref{tab:table_2} to explore the mass-size relation, first proposed by \cite{1981Larson}. \cite{1981Larson} conclude that the mass $M$ contained within a cloud of radius $r$ obeys a power-law of the form:

\begin{equation} \label{eq:mass_size_formula}
M(r) = a \times \left(\frac{r}{\rm pc}\right)^b  
\end{equation}

\noindent \cite{1981Larson} obtain the relation that $M(r) = 460 \;  {\rm M_\odot} \times (\frac{r}{\rm pc})^{1.9}$ or more generally, that the mass of a cloud is proportional to its area, implying constant column density. This law of constant column density has come to be known as one of the fundamental properties of molecular cloud structure \citep{McKee_Ostriker_2007}. Recently, a similar relationship between the masses and lengths $L$ of filaments ($M \propto L^{2}$) has also been found \citep{Hacar_2022} and attributed to turbulent fragmentation. With the goal of testing whether the dimensionality used to define clouds affects our results, we fit the mass-size relation in both 3D and 2D space. Fitting the relation in log-log space, we use a linear-least-squares fitter to obtain $a$, $b$ and their associated uncertainties. For the 3D results we obtain $\log M = 2.9 \times \log r(\pm 0.1) + 0.86(\pm0.3)$ such that:

\begin{equation} \label{eq:3d_eq}
    M(r) = 7 \; \rm M_\odot \times (\frac{\it r}{\rm pc})^{2.9}
\end{equation} 
\\
And for the 2D results, we obtain $\log M = 2.1 \times \log r(\pm 0.2) + 1.59(\pm 0.4)$, such that: 

\begin{equation} \label{eq: 2d_mass_size_equation}
    M(r) = 39 \; \rm M_\odot \times (\frac{\it r}{\rm pc})^{2.1}
\end{equation}

Moreover, we have included in the Appendix (see \S \ref{sec:appendix_a}) another version of the 2D catalog with a minimum extinction threshold when defining the dendrogram of $A_{K_{min}} = 0.1$ mag. This test yielded fewer features, as expected, but maintained a similar mass-size relation of $M(r) = 83 \;  \rm M_\odot \times (\frac{\it r}{\rm pc})^{1.9}$, confirming that the scaling of the mass-size relation does not depend on the threshold used to define cloud boundaries. We also repeat both fits using only the subset of the clouds which are well-studied in the literature (e.g. Perseus, Taurus, Lupus, Chamaeleon, Cepheus, and the Orion complex), finding that the results agree with the full catalog fits within our reported uncertainties.

\begin{figure*} \label{fig:3D_panels}
    \includegraphics[width=1.0\textwidth]{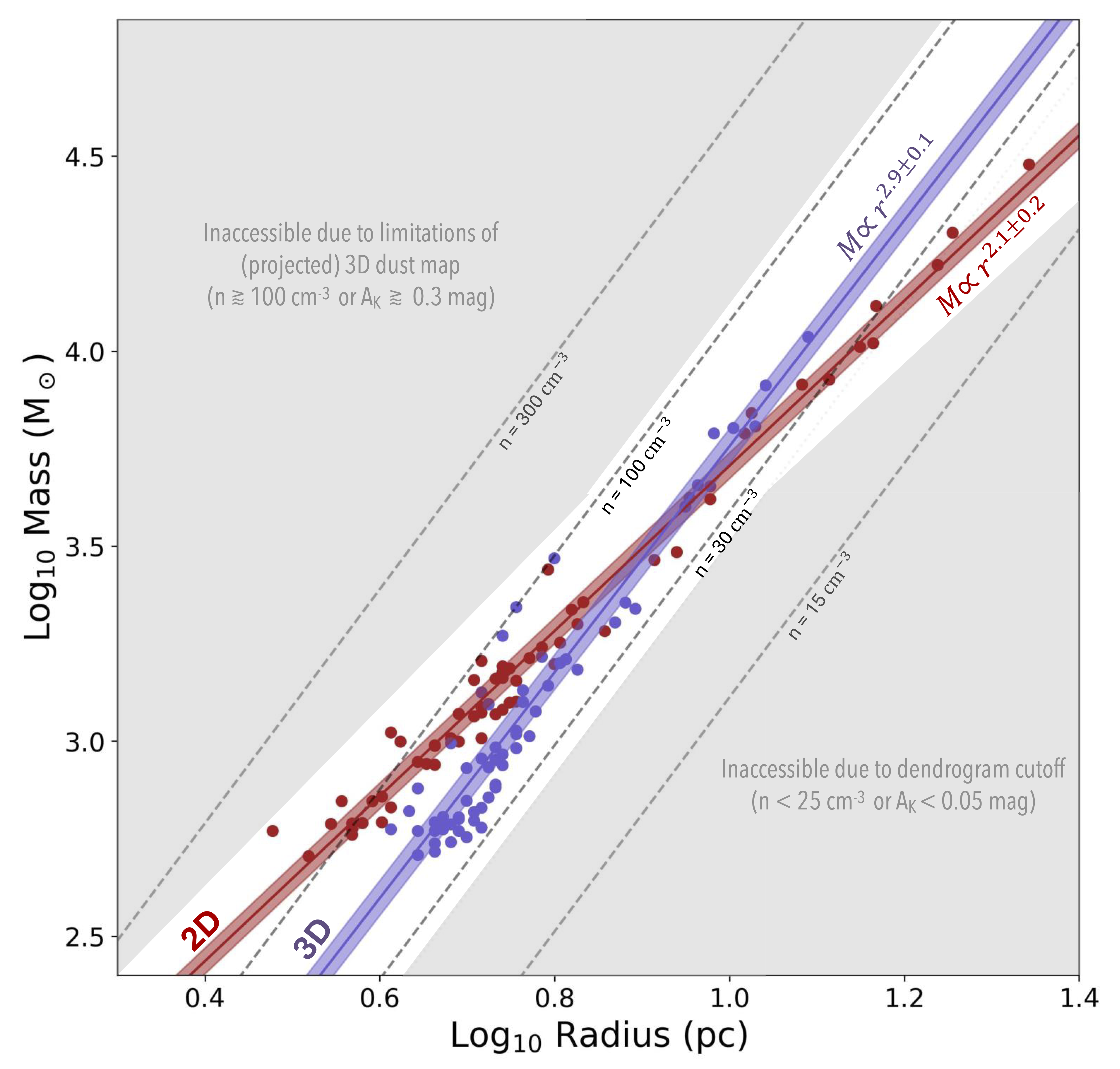}
    \caption{Comparison of the mass-size relation derived from our 3D catalog (blue points) and 2D catalog (red points) of molecular cloud properties. In 3D, $M \propto r^{2.9}$, while in 2D, $M \propto r^{2.1}$. We demarcate the 95\% confidence interval with the thin semi-transparent band around each best-fit line. We include four lines of constant volume density ($n = 15, 30, 100, 300 \rm \;  cm^{-3}$) in dashed semi-transparent black lines. The gray triangles represent the inaccessible volume and column densities, either due to the minimum volume and extinction thresholds required for inclusion in the 3D and 2D dendrograms  ($n < 25 \; \rm cm^{-3}$ and $A_K < 0.05 \; \rm mag$) or the inability of the 3D dust map to probe higher densities due to the map's reliance on optical stellar photometry ($n \gtrsim 100 \; \rm cm^{-3}$ or $A_K \gtrsim 0.2-0.3 \; \rm mag$). }
    \label{fig:mass_size}
\end{figure*}

In Figure \ref{fig:mass_size}, we plot mass versus size for our 3D and 2D catalogs with the best-fits overlaid and four lines of constant volume density ($n = 15, 30, 100, 300 \; \rm cm^{-3} $), assuming purely spherical geometries. Figure \ref{fig:mass_size} shows that the clouds in the 3D catalog lay between $n = 30\; \rm cm^{-3}$ and $n = 100 \; \rm cm^{-3}$ lines of constant volume density, which we will argue in \S \ref{sec:discussion} is a consequence of our dendrogramming procedure and the narrow volume density range probed by the \citet{Leike} 3D dust map. 

\startlongtable
\begin{deluxetable*}{ccccccccccccccc}
\setlength{\tabcolsep}{2pt}
\tabletypesize{\scriptsize}
\tablecaption{3D Cloud Catalog\label{tab:table_1}}
\colnumbers
\tablehead{\colhead{Cloud} & \colhead{Complex} & \colhead{$x$} & \colhead{$y$} & \colhead{$z$} & \colhead{$l$} & \colhead{$b$} & \colhead{$d$} & \colhead{$n_{average}$} & \colhead{$n_{peak}$} &  \colhead{$M$} & \colhead{$r$} & \colhead{$A$} & \colhead{$V$} & \colhead{$\Sigma$} \\
\colhead{} & \colhead{} & \colhead{pc} & \colhead{pc} & \colhead{pc} & \colhead{$^\circ$} & \colhead{$^\circ$} & \colhead{pc} &  \colhead{cm$^{-3}$} & \colhead{cm$^{-3}$} & \colhead{M$_{\odot}$} &  \colhead{pc} & \colhead{pc$^2$} & \colhead{pc$^3$} & \colhead{$\frac{\rm M_\odot}{\rm pc^2}$}}
\startdata
0 & - & -218 & 82 & -157 & 159.2 & -33.9 & 282 & 52 & 314 & 1352 & 5.8 & 106 & 826 & 12 \\
1 & Orion A* & -340 & -236 & -148 & 214.8 & -19.8 & 440 & 38 & 120 & 774 & 5.4 & 90 & 647 & 8 \\
2 & Orion A* & -348 & -184 & -143 & 207.9 & -20.0 & 419 & 45 & 221 & 1624 & 6.5 & 133 & 1163 & 12 \\
3 & Orion A* & -320 & -204 & -135 & 212.6 & -19.6 & 403 & 44 & 1074 & 3996 & 8.9 & 247 & 2923 & 16 \\
4 & - & -288 & -162 & -116 & 209.4 & -19.3 & 350 & 92 & 1349 & 2947 & 6.3 & 123 & 1029 & 23 \\
5 & - & -356 & -158 & -117 & 204.0 & -16.8 & 406 & 53 & 915 & 590 & 4.4 & 61 & 359 & 9 \\
6 & Perseus & -257 & 97 & -102 & 159.3 & -20.4 & 294 & 40 & 154 & 6425 & 10.7 & 361 & 5159 & 17 \\
7 & Orion B* & -351 & -172 & -100 & 206.1 & -14.4 & 404 & 45 & 716 & 10878 & 12.3 & 472 & 7736 & 23 \\
8 & - & 140 & -147 & -113 & 313.5 & -29.1 & 232 & 47 & 427 & 590 & 4.6 & 66 & 406 & 8 \\
9 & Orion $\lambda$* & -363 & -101 & -108 & 195.6 & -16.0 & 392 & 38 & 375 & 2019 & 7.4 & 173 & 1718 & 11 \\
10 & - & -280 & -136 & -102 & 206.0 & -18.2 & 328 & 85 & 1436 & 1867 & 5.5 & 95 & 704 & 19 \\
11 & - & -277 & -114 & -101 & 202.4 & -18.8 & 316 & 90 & 3505 & 2207 & 5.7 & 103 & 790 & 21 \\
12 & Orion  $\lambda$* & -360 & -76 & -86 & 192.0 & -13.2 & 378 & 39 & 838 & 1526 & 6.7 & 140 & 1248 & 10 \\
13 & - & 142 & -145 & -84 & 314.6 & -22.5 & 220 & 66 & 620 & 758 & 4.4 & 62 & 368 & 12 \\
14 & Orion  $\lambda$* & -359 & -129 & -79 & 199.7 & -11.7 & 390 & 39 & 146 & 1031 & 5.9 & 107 & 843 & 9 \\
15 & Orion  $\lambda$* & -350 & -94 & -78 & 195.0 & -12.2 & 371 & 72 & 1321 & 1335 & 5.2 & 85 & 594 & 15 \\
16 & - & -254 & 118 & -76 & 155.0 & -15.3 & 291 & 42 & 364 & 925 & 5.5 & 95 & 697 & 9 \\
17 & - & -61 & -363 & -69 & 260.4 & -10.7 & 374 & 55 & 838 & 1649 & 6.1 & 117 & 953 & 14 \\
18 & - & -129 & -310 & -74 & 247.4 & -12.4 & 344 & 66 & 500 & 595 & 4.1 & 52 & 289 & 11 \\
19 & - & -347 & -140 & -69 & 202.0 & -10.6 & 381 & 36 & 325 & 626 & 5.1 & 82 & 560 & 7 \\
20 & Musca and Chamaeleon & 88 & -153 & -48 & 300.0 & -15.3 & 183 & 53 & 483 & 6162 & 9.6 & 291 & 3734 & 21 \\
21 & - & -212 & -13 & -43 & 183.8 & -11.5 & 216 & 38 & 284 & 660 & 5.1 & 81 & 554 & 8 \\
22 & Taurus & -141 & 20 & -37 & 171.6 & -14.8 & 147 & 47 & 562 & 6352 & 10.1 & 320 & 4312 & 19 \\
23 & - & -41 & -313 & -41 & 262.4 & -7.5 & 318 & 63 & 799 & 1244 & 5.3 & 89 & 632 & 14 \\
24 & - & 191 & 111 & -24 & 30.1 & -6.3 & 223 & 43 & 209 & 548 & 4.6 & 66 & 411 & 8 \\
25 & - & 281 & 101 & -17 & 19.8 & -3.4 & 299 & 37 & 126 & 761 & 5.4 & 91 & 657 & 8 \\
26 & - & -197 & 44 & -15 & 167.3 & -4.4 & 202 & 39 & 214 & 590 & 4.9 & 74 & 488 & 7 \\
27 & - & 272 & 282 & -12 & 46.0 & -1.8 & 392 & 40 & 197 & 4501 & 9.5 & 285 & 3625 & 15 \\
28 & - & 74 & -171 & -13 & 293.4 & -4.0 & 187 & 41 & 275 & 639 & 4.9 & 75 & 495 & 8 \\
29 & - & 215 & 120 & -8 & 29.2 & -2.0 & 247 & 36 & 104 & 718 & 5.3 & 89 & 639 & 8 \\
30 & - & 96 & -159 & -7 & 301.1 & -2.4 & 186 & 43 & 267 & 613 & 4.8 & 72 & 461 & 8 \\
31 & Lupus & 193 & -35 & 12 & 349.6 & 3.5 & 197 & 40 & 263 & 2267 & 7.6 & 180 & 1816 & 12 \\
32 & - & -245 & 112 & 8 & 155.3 & 1.9 & 269 & 44 & 348 & 896 & 5.4 & 90 & 645 & 9 \\
33 & Lupus & 149 & -59 & 16 & 338.5 & 5.9 & 161 & 44 & 398 & 4210 & 9.0 & 253 & 3029 & 16 \\
34 & Pipe and Ophiuchus & 134 & -5 & 29 & 357.8 & 12.3 & 138 & 47 & 432 & 8177 & 11.0 & 378 & 5529 & 21 \\
35 & - & 228 & 114 & 21 & 26.5 & 4.7 & 256 & 35 & 157 & 2188 & 7.8 & 189 & 1969 & 11 \\
36 & - & 201 & 77 & 21 & 21.1 & 5.6 & 217 & 33 & 90 & 601 & 5.2 & 84 & 586 & 7 \\
37 & - & -225 & 263 & 24 & 130.6 & 4.0 & 347 & 39 & 155 & 868 & 5.5 & 95 & 707 & 9 \\
38 & - & -205 & 283 & 24 & 125.9 & 4.0 & 350 & 42 & 162 & 633 & 4.9 & 74 & 485 & 8 \\
39 & - & -148 & 318 & 31 & 115.0 & 5.2 & 352 & 43 & 250 & 1043 & 5.7 & 101 & 772 & 10 \\
40 & - & -175 & 267 & 28 & 123.2 & 5.0 & 321 & 46 & 331 & 511 & 4.4 & 61 & 359 & 8 \\
41 & - & 219 & 113 & 31 & 27.2 & 7.3 & 249 & 39 & 145 & 552 & 4.8 & 71 & 453 & 7 \\
42 & Ophiuchus & 141 & -24 & 38 & 350.0 & 15.1 & 148 & 43 & 180 & 1054 & 5.7 & 102 & 777 & 10 \\
43 & - & -199 & 207 & 39 & 133.9 & 7.9 & 290 & 44 & 420 & 1388 & 6.2 & 121 & 1009 & 11 \\
44 & - & -106 & 218 & 37 & 116.1 & 8.7 & 245 & 42 & 239 & 522 & 4.6 & 65 & 397 & 8 \\
45 & - & -164 & 271 & 39 & 121.2 & 7.1 & 319 & 47 & 584 & 621 & 4.6 & 67 & 420 & 9 \\
46 & Ophiuchus & 107 & 18 & 39 & 9.9 & 19.9 & 116 & 46 & 376 & 640 & 4.7 & 70 & 443 & 9 \\
47 & - & -245 & 221 & 42 & 138.0 & 7.3 & 333 & 44 & 265 & 858 & 5.3 & 88 & 626 & 9 \\
48 & Ophiuchus & 141 & 10 & 43 & 4.1 & 17.2 & 148 & 37 & 298 & 660 & 5.1 & 83 & 570 & 7 \\
49 & Lupus & 138 & -55 & 45 & 338.2 & 16.7 & 156 & 63 & 460 & 663 & 4.3 & 58 & 340 & 11 \\
50 & Lupus & 168 & -71 & 48 & 337.0 & 14.9 & 189 & 43 & 269 & 595 & 4.7 & 70 & 442 & 8 \\
51 & - & -223 & 166 & 48 & 143.3 & 9.9 & 282 & 44 & 185 & 609 & 4.7 & 69 & 439 & 8 \\
52 & - & -132 & 179 & 54 & 126.5 & 13.8 & 230 & 46 & 280 & 1589 & 6.4 & 129 & 1106 & 12 \\
53 & - & -144 & 275 & 56 & 117.7 & 10.4 & 315 & 42 & 434 & 1193 & 6.0 & 112 & 903 & 10 \\
54 & - & -195 & 208 & 64 & 133.2 & 12.6 & 293 & 46 & 340 & 964 & 5.4 & 93 & 675 & 10 \\
55 & - & -191 & 227 & 73 & 130.0 & 13.9 & 306 & 50 & 436 & 904 & 5.2 & 84 & 583 & 10 \\
56 & Cepheus & -78 & 320 & 80 & 103.7 & 13.8 & 339 & 53 & 316 & 854 & 5.0 & 77 & 515 & 11 \\
57 & Cepheus & -133 & 303 & 99 & 113.8 & 16.6 & 346 & 44 & 627 & 4535 & 9.2 & 267 & 3283 & 17 \\
58 & Cepheus & -69 & 318 & 89 & 102.3 & 15.4 & 338 & 68 & 724 & 988 & 4.8 & 73 & 469 & 13 \\
59 & Cepheus & -79 & 328 & 108 & 103.6 & 17.8 & 354 & 35 & 132 & 568 & 5.0 & 78 & 518 & 7 \\
60 & Cepheus & -143 & 293 & 116 & 116.0 & 19.6 & 346 & 44 & 265 & 1065 & 5.7 & 102 & 780 & 10 \\
61 & Cepheus & -128 & 320 & 126 & 111.8 & 20.2 & 367 & 50 & 382 & 1261 & 5.8 & 104 & 802 & 12 \\
62 & - & -165 & 268 & 145 & 121.6 & 24.7 & 347 & 40 & 321 & 960 & 5.7 & 100 & 760 & 9 \\
63 & - & -179 & 265 & 150 & 124.0 & 25.2 & 353 & 43 & 465 & 705 & 5.0 & 78 & 519 & 9 \\
64 & - & -175 & 256 & 181 & 124.4 & 30.2 & 360 & 37 & 115 & 675 & 5.2 & 84 & 580 & 8 \\
\enddata
\tablecomments{Properties of local molecular clouds derived in 3D {\it p-p-p} space. (1) Cloud Index (2) Association with well-studied nearby cloud complexes (3-5) Heliocentric Galactic Cartesian coordinates $x$,$y$,$z$ (6-7) Central Galactic Longitude $l$ and Latitude $b$ (8) Distance (9) Average volume density (10) Peak volume density (11) Mass (12) Radius (13) Surface area (14) Volume (15) Surface mass density.}
\tablecomments{A machine readable version of this table is available at \href{https://doi.org/10.7910/DVN/BFYDG8}{https://doi.org/10.7910/DVN/BFYDG8}.}
\tablenotetext{*}{The Orion Clouds lie at the very edge of the \citet{Leike} 3D dust grid, thus adding more uncertainty to their derived properties and should be treated with caution.}
\end{deluxetable*}

\startlongtable
\begin{deluxetable*}{ccccccccc}
\setlength{\tabcolsep}{2pt}
\tabletypesize{\scriptsize}
\tablecaption{2D Projected Catalog\label{tab:table_2}}
\colnumbers
\tablehead{\colhead{Cloud} & \colhead{Complex} & \colhead{$l$} & \colhead{$b$} & \colhead{$d$} & \colhead{$M$} & \colhead{$r$} & \colhead{$A$} &  \colhead{$\Sigma$} \\
\colhead{} & \colhead{} & \colhead{$^\circ$} & \colhead{$^\circ$} & \colhead{pc} &  \colhead{$\rm M_{\odot}$} & \colhead{pc} & \colhead{pc$^2$} & \colhead{$\rm \frac{\rm M_\odot}{\rm pc^2}$}}
\startdata
0 & - & 159.1 & -33.8 & 282 & 1609 & 5.2 & 84 & 19 \\
1 & Orion A* & 214.9 & -19.6 & 440 & 1184 & 5.2 & 85 & 14 \\
2 & Orion A* & 207.8 & -20.0 & 419 & 2179 & 6.6 & 137 & 16 \\
3 & Orion A* & 212.7 & -19.5 & 403 & 6159 & 10.4 & 341 & 18 \\
4 & - & 209.2 & -19.3 & 350 & 2758 & 6.2 & 121 & 23 \\
6 & Perseus & 159.0 & -20.3 & 294 & 13083 & 14.7 & 679 & 19 \\
$7_0$ & Orion B* & 206.1 & -13.9 & 404 & 20143 & 18.0 & 1013 & 20 \\
$7_1$ & Orion B* & 204.0 & -16.6 & 404 & 595 & 3.7 & 43 & 14 \\
8 & - & 313.5 & -28.9 & 232 & 887 & 4.4 & 59 & 15 \\
9 & Orion $\lambda$* & 195.1 & -16.0 & 392 & 2917 & 8.2 & 209 & 14 \\
12 & Orion $\lambda$* & 192.0 & -13.3 & 378 & 3055 & 8.7 & 238 & 13 \\
13 & - & 313.9 & -22.4 & 220 & 508 & 3.3 & 34 & 15 \\
14 & Orion $\lambda$* & 199.7 & -11.6 & 390 & 1537 & 5.6 & 98 & 16 \\
15 & Orion $\lambda$* & 194.8 & -12.0 & 371 & 1436 & 5.1 & 81 & 18 \\
16 & - & 154.9 & -15.3 & 291 & 1507 & 5.5 & 95 & 16 \\
17 & - & 260.2 & -10.5 & 374 & 1458 & 5.5 & 96 & 15 \\
18 & - & 247.5 & -12.4 & 344 & 590 & 3.0 & 28 & 21 \\
19 & - & 201.9 & -10.4 & 381 & 1635 & 5.9 & 109 & 15 \\
20 & Musca and Chamaeleon & 299.5 & -14.7 & 183 & 10500 & 14.6 & 668 & 16 \\
21 & - & 183.5 & -11.7 & 216 & 1267 & 5.7 & 102 & 12 \\
22 & Taurus & 171.2 & -14.2 & 147 & 16652 & 17.3 & 936 & 18 \\
23 & - & 262.8 & -7.6 & 318 & 615 & 3.5 & 38 & 16 \\
24 & - & 30.4 & -6.0 & 223 & 618 & 3.8 & 46 & 13 \\
25 & - & 20.1 & -3.4 & 299 & 1576 & 6.3 & 124 & 13 \\
26 & - & 167.0 & -4.1 & 202 & 620 & 4.0 & 49 & 13 \\
27 & - & 46.4 & -1.5 & 392 & 10262 & 14.1 & 625 & 16 \\
28 & - & 293.4 & -3.9 & 187 & 975 & 4.6 & 67 & 15 \\
29 & - & 29.3 & -1.8 & 247 & 1206 & 5.5 & 93 & 13 \\
30 & - & 301.3 & -2.3 & 186 & 721 & 4.0 & 49 & 15 \\
31 & Lupus & 349.6 & 3.8 & 197 & 4179 & 9.5 & 285 & 15 \\
32 & - & 155.3 & 1.9 & 269 & 1247 & 5.3 & 88 & 14 \\
33 & Lupus & 338.3 & 6.1 & 161 & 8462 & 13.0 & 530 & 16 \\
34 & Pipe and Ophiuchus & 0.0 & 13.7 & 138 & 30146 & 22.0 & 1516 & 20 \\
35 & - & 26.4 & 5.3 & 256 & 8225 & 12.1 & 463 & 18 \\
36 & - & 21.4 & 5.3 & 217 & 1915 & 7.2 & 161 & 12 \\
37 & - & 131.2 & 4.2 & 347 & 1792 & 6.4 & 130 & 14 \\
38 & - & 126.0 & 4.0 & 350 & 876 & 4.5 & 62 & 14 \\
39 & - & 115.0 & 5.1 & 352 & 1743 & 6.1 & 115 & 15 \\
40 & - & 123.2 & 5.1 & 321 & 577 & 3.7 & 41 & 14 \\
41 & - & 27.3 & 7.3 & 249 & 871 & 4.6 & 66 & 13 \\
42 & Ophiuchus & 350.3 & 15.2 & 148 & 1454 & 5.5 & 95 & 15 \\
43 & - & 133.7 & 7.9 & 290 & 1998 & 6.7 & 142 & 14 \\
44 & - & 115.9 & 8.7 & 245 & 677 & 4.1 & 52 & 13 \\
45 & - & 121.4 & 7.0 & 319 & 616 & 3.7 & 42 & 15 \\
46 & Ophiuchus & 9.5 & 19.9 & 116 & 1018 & 4.8 & 73 & 14 \\
47 & - & 137.9 & 7.5 & 333 & 1158 & 5.1 & 81 & 14 \\
48 & Ophiuchus & 4.5 & 17.3 & 148 & 1172 & 5.4 & 92 & 13 \\
49 & Lupus & 338.4 & 16.9 & 156 & 703 & 3.6 & 39 & 18 \\
51 & - & 143.1 & 10.0 & 282 & 702 & 3.9 & 48 & 15 \\
52 & - & 126.3 & 13.9 & 230 & 2273 & 6.8 & 144 & 16 \\
53 & - & 117.4 & 10.8 & 315 & 1256 & 5.6 & 98 & 13 \\
54 & - & 133.1 & 12.7 & 293 & 1224 & 5.2 & 85 & 14 \\
55 & - & 129.8 & 14.0 & 306 & 1177 & 4.9 & 76 & 16 \\
56 & Cepheus & 103.6 & 13.9 & 339 & 998 & 4.2 & 54 & 18 \\
57 & Cepheus & 113.7 & 16.7 & 346 & 6944 & 10.6 & 353 & 20 \\
58 & Cepheus & 102.2 & 15.4 & 338 & 1055 & 4.1 & 53 & 20 \\
59 & Cepheus & 103.4 & 17.8 & 354 & 1019 & 5.2 & 85 & 12 \\
60 & Cepheus & 116.0 & 19.7 & 346 & 1447 & 5.4 & 92 & 16 \\
61 & Cepheus & 111.7 & 20.2 & 367 & 1556 & 5.5 & 94 & 17 \\
62 & - & 121.4 & 24.8 & 347 & 1431 & 5.7 & 102 & 14 \\
63 & - & 124.0 & 25.3 & 353 & 976 & 4.6 & 65 & 15 \\
64 & - & 124.3 & 30.2 & 360 & 998 & 4.9 & 74 & 14 \\
\enddata
\tablecomments{Properties of the 2D projected local molecular clouds (1)  Cloud index   (2) Association with well-studied nearby cloud complexes (3-4) Central Galactic longitude $l$ and  latitude $b$ (5) Distance (6) Mass (7) Radius (8) Exact area (9) Surface mass density. A small fraction of the 3D clouds did not have a 2D counterpart that met our cloud definition, which accounts for the fact that there are 62 structures identified in 2D after projecting our 65 3D structures from Table \ref{tab:table_1} on to plane of the sky.}
\tablecomments{A machine readable version of this table is available at \href{https://doi.org/10.7910/DVN/BFYDG8}{https://doi.org/10.7910/DVN/BFYDG8}.}
\tablecomments{Projecting feature 7 from 3D into 2D with $A_{K_{min}} = 0.05$ mag yields two components, $7_0$ and $7_1$.}
\tablecomments{A version of this table with a higher minimum threshold for cloud boundary definition, $A_{K,min} = 0.1$ mag, is available in Appendix \S \ref{sec:appendix_a}.}
\tablenotetext{*}{The Orion Clouds lie at the very edge of the \citet{Leike} 3D dust grid, thus adding more uncertainty to their derived properties and should be treated with caution.}
\end{deluxetable*}

\section{Discussion} \label{sec:discussion}

\begin{figure*}[ht!]
\label{catalogs_plotted_together}
    \centering
    \includegraphics[width=1.0\textwidth]{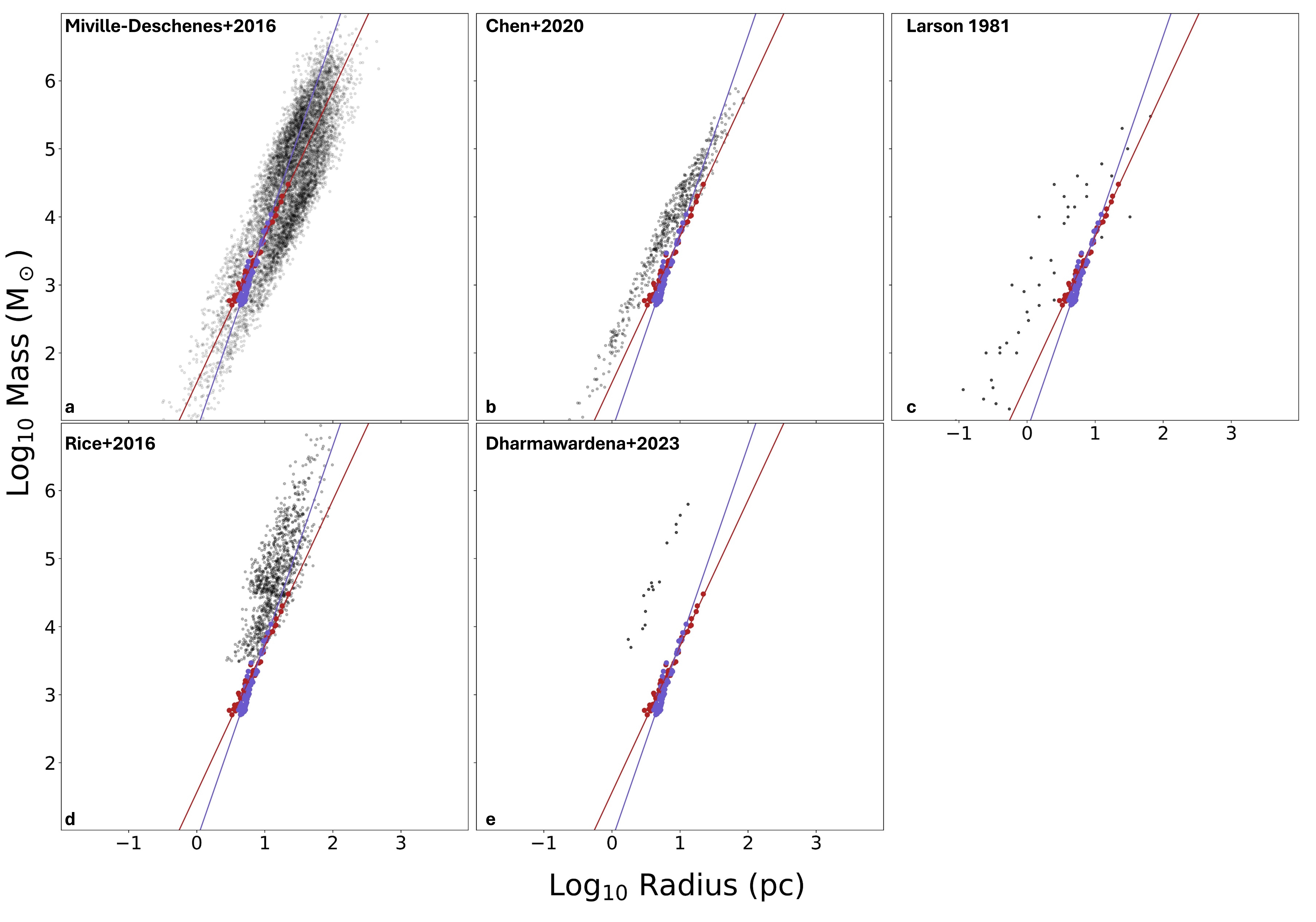}
    \caption{3D and 2D mass-size relations for the 3D dust (blue) and projected 2D dust (red) cloud catalogs, in comparison to five catalogs from the literature, including (a) \citet{Miville_Desch_nes_2016}, (b) \citet{Chen_2020}, (c) \citet{1981Larson}, (d) \citet{2016_Rice} and (e) 
    \citet{Dharmawardena_2022}.}
    \label{fig:all_clouds_way}
\end{figure*}

Here we compare our results for the 3D and 2D mass-size relation with extant results from the literature. We base our comparison on the work of \citet{Lada_2020}, which analyzes both dust-based cloud catalogs and CO-based cloud catalogs to investigate the nature of the mass-size relation in the Milky Way. Specifically, we consider the original cloud sample of \citet{1981Larson}, alongside more recent cloud catalogs re-analyzed in \citet{Lada_2020}, including: the \citet{2016_Rice} catalog based on a dendrogram decomposition of the \citet{Dame_2001} $^{\rm 12}{\rm CO}$ survey; the \citet{Miville_Desch_nes_2016} catalog based on a hierarchical clustering algorithm applied to Gaussian fits of the same \citet{Dame_2001} $^{\rm 12}\rm CO$ survey; and the \citet{Chen_2020} catalog based on the 3D dust map of \citet{Chen_2019}. We also include a comparison to \citet{Dharmawardena_2022}, which is the only other study to extract cloud properties in \textit{p-p-p} space. For consistency, we compare to the \citet{Dharmawardena_2022} primary trunks only catalog (see Table 2 in their work).

 In Figure \ref{fig:all_clouds_way}, we plot our mass-size results in context. Following \citet{Lada_2020}, the catalogs from \citet{2016_Rice}, \citet{Miville_Desch_nes_2016} and  \citet{Chen_2020} are all consistent with $b = 2$ ($M \sim r^{2}$), implying constant surface (column) density across molecular clouds. While some of the clouds in these catalogs may have originally been identified using 3D \textit{position-position-velocity} data \citep{2016_Rice, Miville_Desch_nes_2016} or even 3D \textit{position-position-position} data \citep{Chen_2020}, the masses and/or sizes of these clouds have all been measured in 2D by projecting the 3D data on to the plane of the sky \citep{Chen_2020, 2016_Rice, Miville_Desch_nes_2016}. 
 
By measuring the masses and sizes of fully-resolved molecular clouds \textit{directly} in 3D volume density space, we find a larger power law slope of approximately three, or $b=2.9$. This discrepancy suggests that the power law slope is directly dependent on the number of dimensions used to measure mass and area, with the power law slope consistent with $b=2$ when measuring cloud properties in column density space and $b=3$ when measuring cloud properties in volume density space. We directly test this hypothesis by projecting the 3D volume density data into 2D, re-defining the cloud boundaries in column density space and re-measuring their masses and areas. We find that despite stemming from the same underlying 3D dust data, the projected results are consistent with a shallower power law slope of $b=2.1$. 

\begin{figure*} 
\label{catalogs_plotted_together}
    \centering
    \includegraphics[width=170mm]{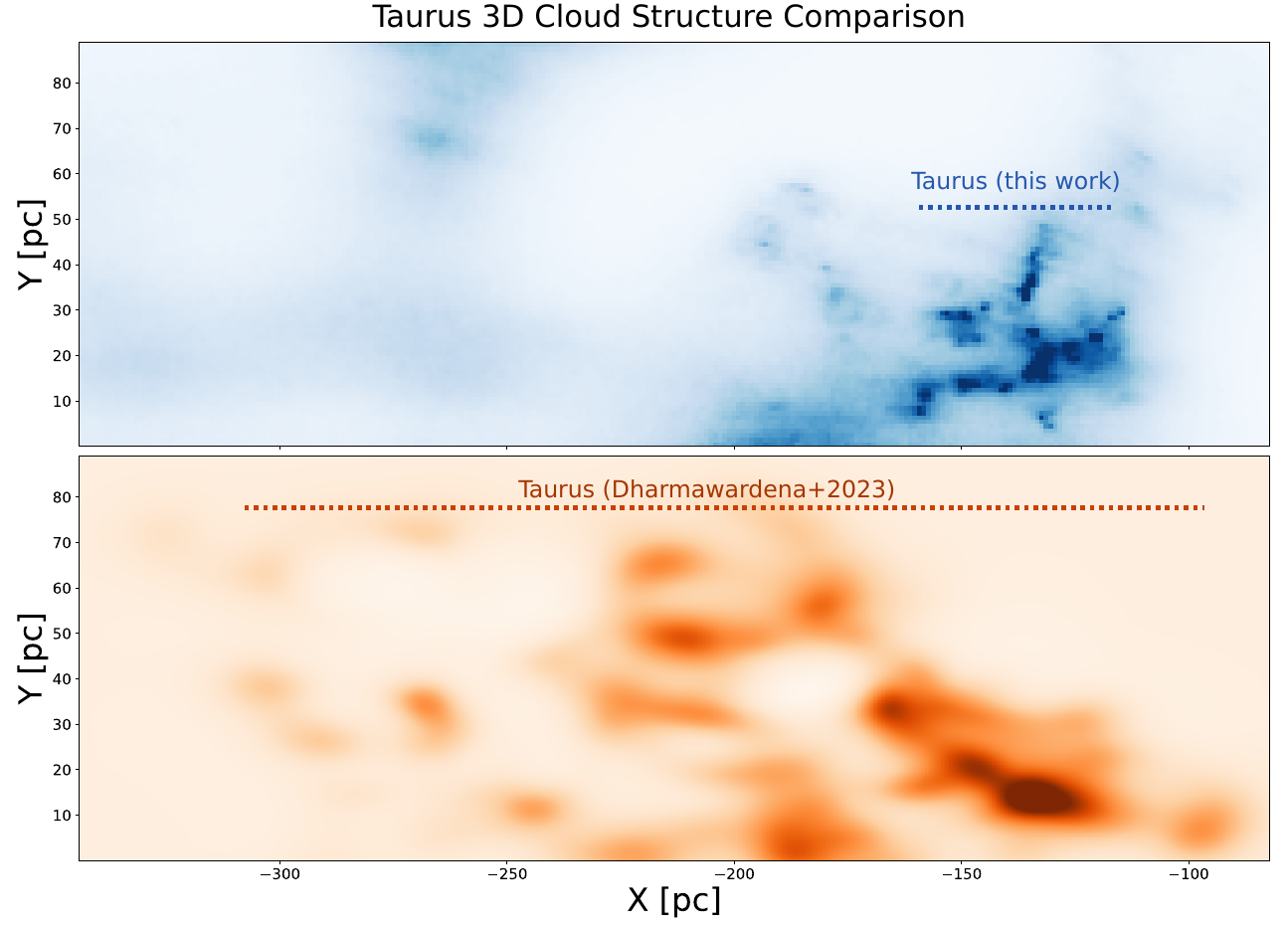}
    \caption{A bird's eye comparison between the Taurus molecular cloud complex as analyzed in \citet{Dharmawardena_2022} (orange) and in this work (blue), based on the 3D dust map from \citet{Leike}. Both 3D dust cutouts have been integrated over the same range in $z$. \citet{Dharmawardena_2022} find roughly an order of magnitude higher cloud masses across their sample of sixteen local clouds compared to this work, likely due to the presence of cloud substructure at farther distances along the line of sight, which is largely not detected in \citet{Leike}.}
    \label{fig:taurus_comp}
\end{figure*}

While \citet{Dharmawardena_2022} argue that their \textit{p-p-p}-based mass-size results are consistent with Larson's $b=1.9$ relationship, Figure \ref{fig:mass_size} suggests that the \citet{Dharmawardena_2022} results may be closer to $b=3$, albeit with an order of magnitude higher cloud masses. We attribute the difference in 3D-derived cloud masses to the existence of more extended cloud substructure along the line of sight in \citet{Dharmawardena_2022}, which is not observed in \citet{Leike}. This claim is supported by Figure \ref{fig:taurus_comp}, which shows that the Taurus molecular cloud --- one of the sixteen cloud complexes in the \citet{Dharmawardena_2022} sample --- spans distances of $d = 93-342$ pc, about a factor of five more extended than we find in this work \citep[$d = 127-170$ pc; based on][]{Leike}.  
 
The higher power law slope for the 3D catalogs observationally validates previous predictions for the scaling of the mass-size relation based on a combination of extant 2D observational results, numerical simulations, and analytic theory. For example, \citet{2010ShettyRosoGoodman} measure masses and sizes of clouds in both volume and column density space based on hydrodynamical simulations, finding $b = 3.03 \pm 0.02$ in 3D and $b=1.95 \pm 0.03$ in 2D. Likewise, \citet{Ballesteros_Paredes_2012} argue that the mass-size relation depends on cloud boundary definition, with column density definitions yielding a power-law slope $b=2$ and volume density definitions yielding $b=3$. The \citet{Ballesteros_Paredes_2012} argument stems from the fact that for clouds with similar boundary definitions, the filling factor of dense structures is small while the filling factor of fluffier structures used to define the cloud boundaries is high, implying that mass should scale with the area in 2D and with volume in 3D \citep[see also e.g.][]{Ballesteros_Paredes_2002, Ballesteros_Paredes_2019}. When clouds are defined as isocontours or isosurfaces above a particular threshold, the average column or volume density of the cloud is similar to the adopted threshold, because a large fraction of pixels or voxels in the cloud lie close to the threshold value. 

As an observational counterpart to the investigations of \citet{2010ShettyRosoGoodman} and \citet{Ballesteros_Paredes_2012}, \citet{Beaumont_2012} examines the mass-size relationship in terms of the column density PDF and its possible variation within and between clouds. Leveraging 2D dust extinction maps from \citet{Lombardi_2010}, \citet{Beaumont_2012} find that for structures defined with a constant extinction threshold, the mean of the column density PDF within each structure varies less than the region-to-region dispersion in area, naturally yielding $M \propto A \propto r^{2}$

In our work, the thresholds used to define cloud boundaries lie close to the minimum volume density ($n_{min}$) or extinction ($A_{K_{min}}$) threshold required for a feature to be included in the dendrogram, and are roughly constant across the sample. 
As seen in Figure \ref{fig:mass_size}, on the lower mass end, the clouds all lie above the $n=25 \; \rm cm^{-3}$ line of constant volume density and $N = 8\times10^{20} \; \rm cm^{-2}$ line of constant column density, which is pre-determined by the minimum volume density ($n_{min}=25 \; \rm cm^{-3}$) and extinction threshold ($A_{K_{min}} = 0.05 \; \rm mag$) required for inclusion in the 3D and 2D catalogs, respectively. On the higher mass end, the clouds lie below $n \approx 100 \; \rm cm^{-3}$ and $N \approx 4\times10^{21} \; \rm cm^{-2}$ which is also pre-determined by the fact that the \citet{Leike} 3D dust map is not sensitive to the densest, most extinguished regions in molecular clouds ($A_K \gtrsim 0.3 \; \rm mag$) due to their reliance on optical photometry \citep[see e.g. discussion in \S 4.4 of][]{Zucker_2021}. Thus, following \citet{Ballesteros_Paredes_2019}, \citet{Beaumont_2012}, and \citet{2010ShettyRosoGoodman}, mass should scale with area in 2D and volume in 3D given the narrow range of column and volume density probed, which we validate here for the first time using the same underlying observational data. 

\section{Conclusions} \label{sec:conclusions}

Using the \cite{Leike} 3D dust map with a distance resolution of 1 pc, we extend the dendrogram technique to \textit{position-position-position} space to extract and measure the properties of clouds in 3D physical space, including their 3D positions, masses, sizes, and volume densities. To compare with extant results, we also create synthetic 2D dust extinction maps from the 3D dust distributions and derive similar properties for the same clouds defined on the plane of the sky. Given the masses and sizes of clouds obtained in 2D and 3D space, we fit the mass-size relation following \citet{1981Larson}. Consistent with predictions from extant observational studies and numerical simulations \citep[see e.g.][]{Beaumont_2012,Ballesteros_Paredes_2019}, we find that our 2D projected mass-size relation, $M \propto r^{2.1}$, agrees with the original \citet{1981Larson} results ($M \propto r^2$), where mass scales according to the cloud's area. However, we obtain a steeper power-law slope for the 3D results, $M \propto r^{2.9}$, where the mass scales according to the cloud's volume. This difference in scaling is a natural consequence of the roughly constant thresholds used to define cloud boundaries, in combination with the fact that the PDF of column and volume density do not systematically scale with structure size. Future work connecting these high-resolution 3D cloud results to complementary tracers of a cloud's CO kinematics should enable further constraints not only on Larson's other relations \citep[see also e.g.][for insight into the Kennicutt-Schmidt relation]{Kainulainen_2021} but also on the physical conditions of star formation within molecular clouds. 
    
\section{Acknowledgements}
The visualization, exploration, and interpretation of data presented in this work were made possible using the glue visualization software, supported under NSF grant numbers OAC-1739657 and CDS\&E:AAG-1908419.  AG and CZ acknowledge support by NASA ADAP grant 80NSSC21K0634 “Knitting Together the Milky Way: An Integrated Model of the Galaxy’s Stars, Gas, and Dust.” CZ acknowledges that support for this work was provided by NASA through the NASA Hubble Fellowship grant \#HST-HF2-51498.001 awarded by the Space Telescope Science Institute, which is operated by the Association of Universities for Research in Astronomy, Inc., for NASA, under contract NAS5-26555. The authors acknowledge Thomas Dame and Andreas Burkert for helpful discussions that contributed to this work. We also thank Thavisha Dharmawardena for help with Figure 6. 

\vspace{5mm}


\software{Astropy \citep{2013A&A...558A..33A,2018AJ....156..123A}, glue \citep{2015_glue}, Unity 3D \citep{unity_soft}, Astrodendro \citep{astrodendro}}





\appendix

\section{Dependence of 2D Catalog Properties on Minimum Extinction Threshold} \label{sec:appendix_a}
To confirm that the 2D mass-size relation is robust to the choice of boundary definition we repeat the cloud extraction procedure described in \S \ref{measuring2D} but using a higher minimum extinction threshold of $A_{K_{min}} = 0.1$ mag (in comparison to $A_{K_{min}} = 0.05$ mag whose results are described in \S \ref{sec:results}). As expected, this higher-extinction-threshold version yields fewer features, as well as divided several of the clouds into multiple components\footnote{If a single 3D feature is broken up into multiple 2D features, the same distance is assigned for each 2D feature.}. Nevertheless, the catalog maintains a similar mass-size relation of $\log M = 1.9 \times \log {r}(\pm 0.2) + 1.92(\pm 0.4)$, such that $M(r) = 83 \;  {\rm M_\odot} \times (\frac{r}{\rm pc})^{1.9}$, confirming that the scaling of the mass-size relation does not depend on the threshold used to define cloud boundaries.
\begin{deluxetable*}{cccccccccc}
\setlength{\tabcolsep}{2pt}
\tabletypesize{\scriptsize}
\tablecaption{2D Catalog Properties with $A_{K_{min}} = 0.1$ mag\label{tab:table_3}}
\colnumbers
\tablehead{\colhead{Cloud} & \colhead{Component ID} & \colhead{Complex} & \colhead{$l$} & \colhead{$b$} & \colhead{$d$} & \colhead{$M$} & \colhead{r} & \colhead{$A$} &  \colhead{$\Sigma$} \\
\colhead{} & \colhead{} & \colhead{} & \colhead{$^\circ$} & \colhead{$^\circ$} & \colhead{ pc} &  \colhead{$\rm M_{\odot}$} & \colhead{pc} & \colhead{$\text{pc}^2$} & \colhead{$\frac{M_\odot}{\rm pc^2}$}}
\startdata
2 & 0 & Orion A* & 207.8 & -20.0 & 419 & 2179 & 6.6 & 137 & 16 \\
4 & 0 & - & 209.2 & -19.3 & 350 & 1879 & 4.4 & 61 & 31 \\
4 & 1 & - & 209.4 & -19.4 & 350 & 1559 & 4.0 & 49 & 32 \\
4 & 2 & - & 209.4 & -19.4 & 350 & 1458 & 3.8 & 46 & 32 \\
4 & 3 & - & 209.4 & -19.4 & 350 & 1356 & 3.7 & 43 & 32 \\
4 & 4 & - & 209.5 & -19.5 & 350 & 1043 & 3.2 & 31 & 34 \\
6 & 0 & Perseus & 159.1 & -20.4 & 294 & 8638 & 10.5 & 347 & 25 \\
7 & 0 & Orion B* & 206.2 & -14.3 & 404 & 13038 & 12.6 & 498 & 26 \\
7 & 1 & Orion B* & 203.9 & -11.8 & 404 & 595 & 2.9 & 26 & 23 \\
11 & 0 & - & 192.0 & -13.3 & 378 & 2868 & 8.5 & 225 & 13 \\
11 & 1 & - & 191.9 & -14.1 & 378 & 1480 & 5.9 & 110 & 14 \\
11 & 2 & - & 192.0 & -11.7 & 378 & 849 & 4.4 & 60 & 14 \\
13 & 0 & - & 199.7 & -11.6 & 390 & 1537 & 5.6 & 98 & 16 \\
15 & 0 & Orion $\lambda$* & 155.2 & -14.7 & 291 & 503 & 2.6 & 21 & 24 \\
16 & 0 & - & 260.2 & -10.5 & 374 & 1458 & 5.5 & 96 & 15 \\
18 & 0 & - & 297.0 & -15.8 & 183 & 1124 & 3.5 & 38 & 30 \\
20 & 0 & Musca and Chamaeleon & 171.2 & -15.1 & 147 & 7596 & 9.4 & 278 & 27 \\
25 & 0 & - & 46.3 & -1.5 & 392 & 9037 & 13.3 & 554 & 16 \\
28 & 0 & - & 301.3 & -2.3 & 186 & 721 & 4.0 & 49 & 15 \\
31 & 0 & Lupus & 338.3 & 6.1 & 161 & 8462 & 13.0 & 530 & 16 \\
32 & 0 & - & 358.4 & 17.5 & 138 & 14493 & 12.9 & 523 & 28 \\
32 & 1 & - & 354.4 & 16.3 & 138 & 4437 & 6.1 & 116 & 38 \\
32 & 2 & - & 357.0 & 19.3 & 138 & 1574 & 3.8 & 44 & 36 \\
33 & 0 & Lupus & 26.4 & 5.3 & 256 & 8225 & 12.1 & 463 & 18 \\
44 & 0 & - & 9.5 & 19.9 & 116 & 1018 & 4.8 & 73 & 14 \\
47 & 0 & - & 338.4 & 16.9 & 156 & 703 & 3.6 & 39 & 18 \\
50 & 0 & Lupus & 126.3 & 13.9 & 230 & 2273 & 6.8 & 144 & 16 \\
53 & 0 & - & 129.8 & 14.0 & 306 & 1177 & 4.9 & 76 & 16 \\
54 & 0 & - & 103.6 & 13.9 & 339 & 998 & 4.2 & 54 & 18 \\
55 & 0 & - & 113.6 & 16.7 & 346 & 4582 & 7.5 & 175 & 26 \\
56 & 0 & Cepheus & 102.2 & 15.4 & 338 & 677 & 2.8 & 24 & 28 \\
59 & 0 & Cepheus & 111.6 & 20.1 & 367 & 657 & 2.8 & 25 & 26 \\
60 & 0 & Cepheus & 124.0 & 25.3 & 347 & 943 & 4.5 & 63 & 15 \\
61 & 0 & Cepheus & 124.0 & 25.3 & 353 & 976 & 4.6 & 65 & 15 \\
\enddata
\tablecomments{Properties of the 2D projected local molecular clouds (1)  Cloud index (2) Component ID (3) Association with well-studied nearby cloud complexes (4-5) Central Galactic longitude $l$ and  latitude $b$ (6) Distance (7) Mass (8) Radius (9) Exact area (10) Surface mass density.}

\tablecomments{A machine readable version of this table is available at \href{https://doi.org/10.7910/DVN/BFYDG8}{https://doi.org/10.7910/DVN/BFYDG8}.}
\tablenotetext{*}{The Orion Clouds lie at the very edge of the \citet{Leike} 3D dust grid, thus adding more uncertainty to their derived properties and should be treated with caution.}
\end{deluxetable*}

\clearpage

\section{Morphological Matching Between 3D Projected and 2D Cloud Features} \label{sec:appendix_b}
\begin{figure}[H]
    \centering
    \text{3D 2D Maps Comparison}\par\medskip
    \includegraphics[width=70mm]{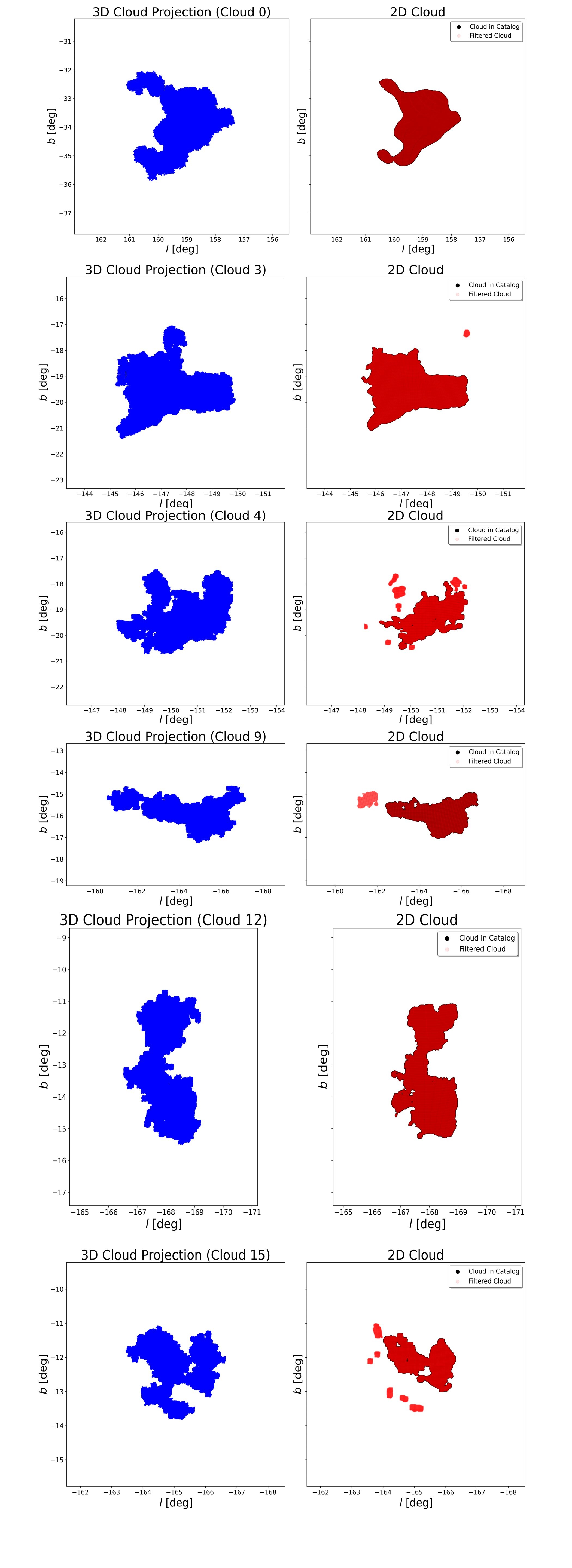}
    \caption{Morphological matching between 3D projected and 2D cloud features for a subset of six clouds in the catalog (Clouds 0, 3, 4, 9, 12, 15). The semi-transparent red on the 2D cloud panels denote 2D projected cloud components that are filtered out from the catalog, since they did not meet our minimum mass or radius requirements.}
    \label{fig:morphological_matching}
\end{figure}

After projecting the 3D data on to the plane of the sky to derive the 2D Cloud features, we compare the morphological agreement between each 3D projected feature and its corresponding 2D cloud counterpart. As seen in Figure \ref{fig:morphological_matching} for a subset of the sample, we overall find good morphological agreement between the 3D and 2D clouds. The remainder of the morphological maps can be accessed at \href{ https://doi.org/10.7910/DVN/BFYDG8}{ https://doi.org/10.7910/DVN/BFYDG8}.

\newpage

\bibliography{References.bib}
\bibliographystyle{aasjournal}

\end{document}